\documentclass[11pt]{article}
\usepackage{latexsym}
\usepackage{amsmath} % it subsumes: amstext, amsbsy, amsopn
\usepackage{amscd}     \usepackage{amsxtra}
\usepackage{upref}     \usepackage{amsthm}
\usepackage{amssymb}   %\usepackage{mathrsfs}
\usepackage{amsfonts}
\begin{document}

\begin{center}
{\bf ON THE INFLUENCE OF THE UNIT CELL TYPE ON THE WAVE \\FUNCTION IN
ONE-DIMENSIONAL KRONIG-PENNEY MODELS}
\end{center}
\begin{center}
%\author{
{\bf M. G. Tashkova$^1$, A. M. Miteva$^2$ and S. G.
Donev}$^{3}$$^($\footnote{e-mail: sdonev@inrne.bas.bg}$^)$
\end{center}
%\date{}
%\maketitle
\vskip 0.4cm
\hspace{1cm}$^1$Institute of Organic Chemistry with Center of
Phytochemistry, Bulgarian \\ \hspace*{1.2cm}Academy of Sciences, Acad.
G.Bonchev Str.  9, 1113 Sofia, Bulgaria
\vskip 0.3cm
\hspace{0.4cm}$^2$Space Research Institute,
Bulgarian Academy of Sciences, P.O.Box 799,\\ \hspace*{1.2cm}1000 Sofia,
Bulgaria
\vskip 0.3cm
\hspace{0.4cm}$^3$Institute for Nuclear Research and Nuclear
Energy, Bulgarian Academy of\\ \hspace*{1.2cm}Sciences,
blvd.Tzarigradsko chaussee 72, 1784 Sofia, Bulgaria
\vskip 1cm
%\date{}
%\maketitle



\vskip 0.8cm

\noindent
{\bf Abstract}.- We study the influence of the different choice of unit cells
on the Bloch solutions of Schr\"odinger equation for one-dimensional
periodic Kronig-Penney models with rectangular potential barriers or
potential wells and partially constant effective mass. We obtain generalized
Kronig-Penney relations for bulk states and exact expressions for the
corresponding periodic parts of the Block wave functions for the two possible
choices of the matching conditions in an unit cell. We show also that our
analytic expressions reduce to the well known expressions for the
Kronig-Penney relations and Bloch waves under constant effective mass and
appropriate matching conditions

\newpage
The superlattice (SL) electronic structure is of primary importance and
interest in both fundamental and applied aspects, and it has been extensively
studied recently [1-10]. Theoretical studies of the electronic properties of
SLs may be carried out in the framework of one-dimensional Kronig-Penney (KP)
models [4,11], in particular, calculating the electronic surface states
[2,4-6], the electronic interface states [3,7-9] and the electronic bulk
states [10] in the semiconductor superlattices. The basic point in these
considerations is to determine the general solution to the corresponding
one-dimensional Schr\"odinger equation together with the conditions for
existence of surface and interface states, which conditions depend strongly
on the corresponding wave functions [2-9]. It hhas been noted by several
authors [12-14] that the kind of the wave function depends on the choice of
the unit cell, therefore, it seems very important to have a more detailed
knowledge of this dependence between the choice of the unit cell and the wave
function.

The aim of this letter is to consider some aspects of this problem, in
particular, how the periodic parts of the Bloch solutions to the
corresponding Schr\"odinger equation with periodic potential and partially
constant effective mass depend on the choice of the unit cell, more precizely,
on the matching points for the wave functions and their derivatives in this
unit cell.

We are going to consider two cases of infinite one-dimensional KP-models:
with rectangular potential barriers given on Fig.1, and with rectangular
potential wells given on Fig.2. In the first case (potential barriers) the
potential $V(x)$, determined by the conditions:
\begin{equation}
V(x)=\
\begin{cases} V=const, & \Omega^1(x): nb+(n+1)a\le x\le (n+1)(a+b),\\
		 0, &\Omega^2(x): n(a+b)\le x\le nb+(n+1)a,\ \ \       %1%
(n=0,\pm1,\pm2,\dots)
\end{cases}
\end{equation}
is periodic with respect to $x$ with a period equal to the lattice constant
$(a+b)$, i.e. $V(x)=V(x+(a+b))$. For an electron with a partially constant
effective mass $m(x)$
\begin{equation}
m(x)=
\begin{cases} m_1=const, &\Omega^1(x): nb+(n+1)a\le x\le (n+1)(a+b)\\
	      m_2=const, &\Omega^2(x): n(a+b)\le x\le nb+(n+1)a,\ \ \    %2%
(n=0,\pm1,\pm2,\dots)
\end{cases}
\end{equation}
moving in the periodic potential field (1), the one-dimensional stationary
Schr\"odinger equation
\begin{equation}
\frac{d^2\Psi(x)}{dx^2}+
\frac{2m(x)}{\hbar^2}\left[E-V(x)\right]\Psi(x)=0   %3%
\end{equation}
has, according to the Bloch-Floquet theorem [4,11], the
following periodic in $x$ solutions with period $(a+b)$:
\begin{equation}
\psi_I(x)=A\,exp\,\left[(-i\alpha +\gamma)x\right]+            %4%
B\,exp\,\left[(-i\alpha-\gamma)x\right]
\end{equation}
in "regions I": $\Omega^1(x): nb+(n+1)a\le x\le (n+1)(a+b)$, and of the form
\begin{equation}
\psi_{II}(x)=C\,exp\,\left[i(-\alpha +\beta)x\right]+
D\,exp\,\left[i(-\alpha-\beta)x\right]
\end{equation}
in "regions II": $\Omega^2(x): n(a+b)\le x\le nb+(n+1)a$. Further in our
consideration we assume the quantities $\beta$ and $\gamma$, introduced by
(4) and (5), to be real and positive and their explicit dependences on the
electron energy $E, (0<E<V)$, are given by
\begin{equation}
\beta=\sqrt{\frac{2m_2}{\hbar^2}E},\quad              %6%
\gamma=\sqrt{\frac{2m_1}{\hbar^2}(V-E)},
\end{equation}
and $\alpha$ is determined by the matching conditions. To obtain the
coefficients $A,B,C$ and $D$ of the wave functions (4) and (5)  we make use
of the matching conditions at three successive points $(x_1,x_2,x_3)$ in one
unit cell such, that $|x_1-x_3|=(a+b)$, and $|x_1-x_2|=a,\ \ |x_2-x_3|=b$, or
$|x_1-x_2|=b,\ \ |x_2-x_3|=a$, and we assume $x_1\le x_2\le x_3$ (Fig.1).
Since there exist only two ways to choose these three points on the $x$
axis we introduce them by the following relations:
\begin{align}
x_1+a &=x_2,\quad x_2+b=x_3,\\                %7%
x_1+b &=x_2,\quad x_2+a=x_3,                  %8%
\end{align}
Relations (7) and (8) specify the two types of unit cells: the unit cell of
type "well-barrier" which we call {\it KP-1 type cell} (Fig.1a), and the unit
cell of type "barrier-well", called {\it KP-2 type cell} (Fig.1b),
respectively.  Thus, if we use the matching conditions used in [1,3,15-17]
and denote $y=\frac{m_2}{m_1}$ , the solutions (4) and (5) for the unit cells
(7) and (8) must obey the following conditions  at $x=x_2$:
\begin{gather}
\begin{split}
\psi_I(x_2) &=\psi_{II}(x_2)\\                               %9%
y\frac{d\psi_I}{dx}(x_2) &=\frac{d\psi_{II}}{dx}(x_2).
\end{split}
\end{gather}
It follows from the periodicity of $\psi(x)$ that the solutions (4) and (5)
and their first derivatives must satisfy also at the points $x=x_1$ and
$x=x_3$ the conditions
\begin{gather}
\begin{split}
\psi_I(x_3)&=\psi_{II}(x_1)\\
y\frac{d\psi_I}{dx}(x_3)&=\frac{d\psi_{II}}{dx}(x_1)          %10%
\end{split}
\end{gather}
for the KP-1 type unit cell defined by (7), and the boundary conditions
\begin{gather}
\begin{split}
\psi_I(x_1)&=\psi_{II}(x_3)\\
y\frac{d\psi_I}{dx}(x_1)&=\frac{d\psi_{II}}{dx}(x_3)
\end{split}
\end{gather}
for the KP-2 type unit cell defined by (8). From (9) and (10) and from (9)
and (11) we obtain two systems of four linear homogeneous equations each, for
the four constants  $A,B,C$ and $D$, reflecting the differences of the two
types unit cells. These two systems have nontrivial solutions only if the
corresponding determinants are equal to zero, which leads to the same
equation for $\beta$ and $\gamma$ given by
\begin{equation}
cos\left[\alpha(a+b)\right]=ch(\gamma b)cos(\beta a)+
\frac{(1-y)^2\alpha^2+y^2\gamma^2 -\beta^2}{2y\beta\gamma}    %12%
sh(\gamma b)sin(\beta a).
\end{equation}
This equation (12) we call {\it generalized Kronig-Penney relation}.

Since only three of the four coefficients $A,B,C$ and $D$ are linearly
independent we can express any three of them through the fourth. So, when
$A=\mu_I^jB$ and $C=\lambda_{II}^jD$, where the index $j=1,2$ denotes the two
types unit cells, the corresponding periodic wave functions (4) and (5)
become
\begin{equation}
\psi_I^j(x)=B\left\{\mu_I^j\,exp\left[(-i\alpha +\gamma)x\right]+
exp\left[(-i\alpha -\gamma)x\right]\right\},\quad (j=1,2),
\end{equation}
\begin{equation}
\psi_{II}^j(x)=D\left\{\lambda_{II}^j\,exp\left[i(-\alpha
+\beta)x\right]+
exp\left[i(-\alpha -\beta)x\right]\right\},\quad (j=1,2).
\end{equation}
For the KP-1 unit cell the corresponding nonhomogeneous system of equations
(9) and (10) gives the following expression for $\mu^1_I$:
\begin{equation}
\mu^1_I=\frac{cos(\beta a)-\left[i(1-y)\frac{\alpha}{\beta}
-y\frac{\gamma}{\beta}\right]sin(\beta a)-exp\left[-i\alpha (a+b)-\gamma
b\right]} {-cos(\beta a)+\left[i(1-y)\frac\alpha\beta
+y\frac\gamma\beta\right]sin(\beta a)
+exp\left[-i\alpha (a+b)+\gamma b\right]}
exp(-2\gamma x_2).                                        %15%
\end{equation}
Similarly, the nonhomogeneous system of equations
(9) and (10) gives for $\lambda^1_{II}$:
\begin{equation}
\lambda^1_{II}=\frac{ch(\gamma b)-
i\left[\left(\frac{1-y}{y}\right)\frac{\alpha}{\gamma}
+\left(\frac 1y\right)\frac{\beta}{\gamma}\right]sh(\gamma
b)-exp\left\{i\left[\alpha
(a+b)+\beta a\right]\right\}}
{-ch(\gamma b)+i\left[\left(\frac{1-y}{y}\right)\frac\alpha\gamma
-\left(\frac 1y\right)\frac\beta\gamma\right]sh(\gamma
b)+exp\left\{i\left[\alpha
(a+b)-\beta a\right]\right\}} exp(-i2\beta x_2).                   %16%
\end{equation}
In the same way for the KP-2 unit cell, from the corresponding nonhomogeneous
system of equations (9) and (11) for the coefficients $\mu_I^2$ and
$\lambda_{II}^2$ we obtain
\begin{equation}
\mu^2_I=\frac{cos(\beta a)+\left[i(1-y)\frac{\alpha}{\beta}
-y\frac{\gamma}{\beta}\right]sin(\beta a)-exp\left[i\alpha (a+b)+\gamma
b\right]}
{-cos(\beta a)-\left[i(1-y)\frac\alpha\beta
+y\frac\gamma\beta\right]sin(\beta a)+exp\left[i\alpha (a+b)-\gamma b\right]}
exp(-2\gamma x_2),
\end{equation}
\begin{equation}
\lambda^2_{II}=\frac{ch(\gamma b)+
i\left[\left(\frac{1-y}{y}\right)\frac{\alpha}{\gamma}
+\left(\frac 1y\right)\frac{\beta}{\gamma}\right]sh(\gamma
b)-exp\left\{-i\left[\alpha (a+b)+\beta a\right]\right\}} {-ch(\gamma
b)-i\left[\left(\frac{1-y}{y}\right)\frac\alpha\gamma -\left(\frac
1y\right)\frac\beta\gamma\right]sh(\gamma b)+exp\left\{-i\left[\alpha
(a+b)-\beta a\right]\right\}} exp(-i2\beta x_2).          %18%
\end{equation}
Hence, for the both
types of unit cells (7) and (8) it is seen from (15) and (17) that the wave
functions (13) in the "region I" are different:  $\psi^1_I\neq \psi^2_I$; and
from (16) and (18) we see that the wave functions (14) in the "region II" are
also different: $\psi^1_{II}\neq \psi^2_{II}$. Consequently, the choice of
matching points inside the unit cell through relations (7)-(11) determines
the form of the periodic parts (13) and (14) of Bloch solutions for the
model under consideration.

Let's consider the case $y=1$, i.e. when the electron effective mass (2) is
constant: $m_1=m_2=m_o$. In this case $\beta$ and $\gamma$ in (6) become
\begin{equation}
\beta_o=\sqrt{\frac{2m_o}{\hbar^2}E},\quad
\gamma_o=\sqrt{\frac{2m_o}{\hbar^2}(V-E)}                     %19%
\end{equation}
and equation (12) is reduced to the well-known Kronig-Penney equation [11].
If for the KP-1 model (Fig.1a) we set $x_2=a$ in relation (7), and then,
making use of (16) and (19), we find for the coefficient
$\lambda^I_{II}\equiv\left(\lambda^I_{II}\right)^B$ the following value:
\begin{equation}
\left(\lambda^I_{II}\right)^B=
\frac{ch(\gamma_ob)-i\frac{\beta_o}{\gamma_o}sh(\gamma_ob)-
exp\left\{i\left[\alpha(a+b)+\beta_oa\right]\right\}}            %20%
{-ch(\gamma_ob)-i\frac{\beta_o}{\gamma_o}sh(\gamma_ob)+
exp\left\{i\left[\alpha(a+b)-\beta_oa\right]\right\}}
exp(-i2\beta_oa),
\end{equation}
which coincides with the corresponding "expression (17)" in Bloss' paper [2].
For the KP-2 model (Fig.1b) if $x_2=0$ in (8) we have the classical KP-model
[11]. In this case from (17) and (19) and from (18) and (19) we find the
following expressions for the corresponding coefficients $\mu_I^2\equiv
\left(\mu_I^2\right)^{KP}$ and
$\lambda_{II}^2\equiv\left(\lambda_{II}^2\right)^{KP}$:
\begin{equation}
(\mu_I^2)^{KP}=
\frac{cos(\beta_oa)-\frac{\gamma_o}{\beta_o}
sin(\beta_oa)-exp[i\alpha(a+b)+\gamma_ob]}
{-cos(\beta_oa)-\frac{\gamma_o}{\beta_o}                         %21%
sin(\beta_oa)+exp[i\alpha(a+b)-\gamma_ob]}
\end{equation}
\begin{equation}
(\lambda_{II}^2)^{KP}=
\frac{ch(\gamma_ob)+i\frac{\beta_o}{\gamma_o}
sh(\gamma_ob)-exp\left\{-i[\alpha(a+b)+\beta_oa]\right\}}        %22%
{-ch(\gamma_ob)+i\frac{\beta_o}{\gamma_o}
sh(\gamma_ob)+exp\left\{-i[\alpha(a+b)-\beta_oa]\right\}}
\end{equation}
These values (21) of $(\mu_I^2)^{KP}$ and (22) of $(\lambda_{II}^2)^{KP}$
have been implicitly used in [11]. However, choosing $x_2=\frac12 b$ in (8),
from (17) and (19) and from (18) and (19) we find that the corresponding
coefficients  $\mu_I^2\equiv (\mu_I^2)^G$ and
$\lambda_{II}^2\equiv (\lambda_{II}^2)^G$, given by
\begin{equation}
(\mu_I^2)^{G}=
\frac{cos(\beta_oa)-\frac{\gamma_o}{\beta_o}
sin(\beta_oa)-exp[i\alpha(a+b)+\gamma_ob]}                  %23%
{-cos(\beta_oa)-\frac{\gamma_o}{\beta_o}
sin(\beta_oa)+exp[i\alpha(a+b)-\gamma_ob]}exp(-\gamma_ob)
\end{equation}
\begin{equation}
(\lambda_{II}^2)^{G}=
\frac{ch(\gamma_ob)+i\frac{\beta_o}{\gamma_o}
sh(\gamma_ob)-exp\left\{-i[\alpha(a+b)+\beta_oa]\right\}}        %24%
{-ch(\gamma_ob)+i\frac{\beta_o}{\gamma_o}
sh(\gamma_ob)+exp\left\{-i[\alpha(a+b)-\beta_oa]\right\}}exp(-i\beta_ob)
\end{equation}
are identical to Gubanov's results [12]. Comparing (21) with (23), and (22)
with (24), we immediately obtain the following relations between them:
\[
(\mu_I^2)^G=(\mu_I^2)^{KP}exp(-\gamma_ob),\quad
(\lambda_{II}^2)^G=(\lambda_{II}^2)^{KP}exp(-i\beta_ob).
\]
Note that the exponential factors in (23) and (24) is due only to the
different choice of the coordinate origin in the unit cell.

The case of the Kronig-Penney models with rectangular potential wells and
partially constant effective mass (2) we treat in the same way, and the
corresponding results are obtained taking into account the periodic potential
$V(x)$ (Fig.2):
\begin{equation}
V(x)=
\begin{cases} 0,  &\Omega^1(x): n(a+b)\le x\le nb +(n+1)a\\
	      -V, &\Omega^2(x): nb+(n+1)a\le x\le (n+1)(a+b),\ \ \  %25%
(n=0,\pm1,\pm2,\dots)
\end{cases}
\end{equation}
where $V>0$. Relations (7) and (8) again specify two types of unit cells: the
unit cell of the type "well-barrier" called {\it KP-1a type cell} (Fig.2a),
and the unit cell of the type "barrier-well" called {\it KP-2a type cell}
(Fig.2b). In the regions with constant potentials the quantities $\theta$ and
$\varphi$, which now play the roles of $\gamma$ and $\beta$ in (4) and (5),
are related to the electron energy $E$ by means of the equations
\[
\theta^2=\frac{2m_1}{\hbar^2} E,\quad
\varphi^2=\frac{2m_2}{\hbar^2}(E+V).
\]
In the case $\theta^2>0,\ \varphi^2>0$ the Schr\"odinger equation (3)
for an electron with effective mass (2), moving in the potential field (25),
for the both types of unit cells has nontrivial solutions only if the
quantities $\theta$ and $\varphi$ satisfy the corresponding {\it generalized
Kronig-Penney relation}
\begin{equation}
cos\left[\alpha(a+b)\right]=cos(\theta a)cos(\varphi b)+           %26%
\frac{(1-y)^2\alpha^2-(y^2\theta^2 +\varphi^2)}{2y\theta\varphi}
sin(\theta a)sin(\varphi b),
\end{equation}
where $y=\frac{m_2}{m_1}$. Now the periodic part $\phi(x)$ of the Bloch wave
function is of the form
\begin{equation}
\phi^j_I(x)=B\left\{\nu^j_I exp[i(-\alpha
+\theta)x]+exp[i(-\alpha-\theta)x]\right\},\ \ (j=1,2),          %27%
\end{equation}
in "regions I" (barriers): $\Omega^1(x): n(a+b)\le x\le nb +(n+1)a$, and of
the form
\begin{equation}
\phi^j_{II}(x)=D\left\{\chi^j_{II} exp[i(-\alpha               %28%
+\varphi)x]+exp[i(-\alpha-\varphi)x]\right\},\ \ (j=1,2)
\end{equation}
in "regions II" (wells): $\Omega^2(x): nb+(n+1)a\le x\le (n+1)(a+b)$. For the
KP-1a unit cell, determined by condition (7), the corresponding
nonhomogeneous system of equations (9) and (10) gives the following
expressions for $\nu^1_I$ and $\chi^1_{II}$:
\begin{equation}
\nu^1_I=\frac{cos(\varphi b)-i\left[(1-y)\frac{\alpha}{\varphi}
-y\frac{\theta}{\varphi}\right]sin(\varphi b)-
exp\left\{-i\left[\alpha (a+b)+\theta a\right]\right\}}        %29%
{-cos(\varphi b)+i\left[(1-y)\frac{\alpha}{\varphi}
+y\frac{\theta}{\varphi}\right]sin(\varphi b)+
exp\left\{-i\left[\alpha (a+b)-\theta a\right]\right\}} exp(-i2\theta x_2),
\end{equation}
\begin{equation}
\chi^1_{II}=\frac{cos(\theta
a)-i\left[\left(\frac{1-y}{y}\right)\frac{\alpha}{\theta}
+\left(\frac1y\right)\frac{\varphi}{\theta}\right]sin(\theta a)-
exp\left\{i\left[\alpha (a+b)+\varphi b\right]\right\}}
{-cos(\theta a)+i\left[\left(\frac{1-y}{y}\right)\frac{\alpha}{\theta}
-\left(\frac1y\right)\frac{\varphi}{\theta}\right]sin(\theta a)+      %30%
exp\left\{i\left[\alpha (a+b)-\varphi b\right]\right\}}
exp(-i2\varphi x_2),
\end{equation}
For the KP-2a unit cell, determined by (8), the coefficients $\nu_I^2$ and
$\chi_{II}^2$ are obtained from the nonhomogeneous system of equations (9) and
(11) are given by:
\begin{equation}
\nu^2_I=\frac{cos(\varphi b)+i\left[(1-y)\frac{\alpha}{\varphi}
-y\frac{\theta}{\varphi}\right]sin(\varphi b)-
exp\left\{i\left[\alpha (a+b)+\theta a\right]\right\}}        %31%
{-cos(\varphi b)-i\left[(1-y)\frac{\alpha}{\varphi}
+y\frac{\theta}{\varphi}\right]sin(\varphi b)+
exp\left\{i\left[\alpha (a+b)-\theta a\right]\right\}} exp(-i2\theta x_2),
\end{equation}
\begin{equation}
\chi^2_{II}=\frac
{cos(\theta a)+i\left[\left(\frac{1-y}{y}\right)\frac{\alpha}{\theta}
+\left(\frac1y\right)\frac{\varphi}{\theta}\right]sin(\theta a)-
exp\left\{-i\left[\alpha (a+b)+\varphi b\right]\right\}}
{-cos(\theta a)-i\left[\left(\frac{1-y}{y}\right)\frac{\alpha}{\theta}
-\left(\frac1y\right)\frac{\varphi}{\theta}\right]sin(\theta a)+      %32%
exp\left\{-i\left[\alpha (a+b)-\varphi b\right]\right\}}
exp(-i2\varphi x_2).
\end{equation}
It is clearly seen that the corresponding wave functions (27) and (28)
satisfy the inequalities $\phi_I^1\neq\phi^2_I$ and
$\phi^1_{II}\neq\phi^2_{II}$.

If the electron energy is negative: $E=-\varepsilon, \varepsilon>0$ and
$-V<E<0$, we have $\theta^2=-\frac{2m}{\hbar^2}\varepsilon<0$ and
$\varphi^2=\frac{2m_2}{\hbar^2}(E+V)>0$, and denoting
$\frac{2m}{\hbar^2}\varepsilon\equiv k^2$,
the generalized Kronig-Penney relation (26) becomes
\begin{equation}
cos\left[\alpha(a+b)\right]=ch(ka)cos(\varphi b)+           %33%
\frac{(1-y)^2\alpha^2-(y^2 k^2 +\varphi^2)}{2y\varphi k}
sh(ka)sin(\varphi b).
\end{equation}
In this case the quantity $\gamma^1_I$ (29) for the
KP-1a model becomes
\begin{equation}
\left(\nu^1_I\right)^{\pm}=\frac{cos(\varphi
b)-\left[i(1-y)\frac{\alpha}{\varphi}
\pm y\frac{k}{\varphi}\right]sin(\varphi b)-
exp\left[-i\alpha (a+b)\pm ka\right]}                       %34%
{-cos(\varphi b)+\left[i(1-y)\frac{\alpha}{\varphi}
\mp y\frac{k}{\varphi}\right]sin(\varphi b)+
exp\left[-i\alpha (a+b)\mp ka\right]} exp(\pm 2kx_2),
\end{equation}
and from (30) we obtain also
$\left(\chi^1_{II}\right)^{+}=\left(\chi^1_{II}\right)^{-}=
\left(\chi^1_{II}\right)^{\pm}$
\begin{equation}
\left(\chi^1_{II}\right)^{\pm}=\frac
{ch(ka)-i\left[\left(\frac{1-y}{y}\right)\frac{\alpha}{k}
+\left(\frac1y\right)\frac{\varphi}{k}\right]sh(ka)-
exp\left\{i\left[\alpha (a+b)+\varphi b\right]\right\}}
{-ch(ka)+i\left[\left(\frac{1-y}{y}\right)\frac{\alpha}{k}
-\left(\frac1y\right)\frac{\varphi}{k}\right]sh(ka)+      %35%
exp\left\{i\left[\alpha (a+b)-\varphi b\right]\right\}}
exp(-i2\varphi x_2).
\end{equation}
For the KP-2a model the quantity $\nu^2_I$ (31) becomes
\begin{equation}
\left(\nu^2_I\right)^{\pm}=\frac
{cos(\varphi b)+\left[i(1-y)\frac{\alpha}{\varphi}
\pm y\frac{k}{\varphi}\right]sin(\varphi b)-
exp\left[i\alpha (a+b)\mp ka\right]}                      %36%
{-cos(\varphi b)-\left[i(1-y)\frac{\alpha}{\varphi}
\mp y\frac{k}{\varphi}\right]sin(\varphi b)+
exp\left[i\alpha (a+b)\pm ka\right]} exp(\pm 2kx_2),
\end{equation}
and from (32) we again have $\left(\chi^2_{II}\right)^{+}=
\left(\chi^2_{II}\right)^{-}=\left(\chi^2_{II}\right)^{\pm}$:
\begin{equation}
\left(\chi^2_{II}\right)^{\pm}=\frac
{ch(ka)+i\left[\left(\frac{1-y}{y}\right)\frac{\alpha}{k}
+\left(\frac1y\right)\frac{\varphi}{k}\right]sh(ka)-
exp\left\{-i\left[\alpha (a+b)+\varphi b\right]\right\}}
{-ch(ka)-i\left[\left(\frac{1-y}{y}\right)\frac{\alpha}{k}
-\left(\frac1y\right)\frac{\varphi}{k}\right]sh(ka)+      %37%
exp\left\{-i\left[\alpha (a+b)-\varphi b\right]\right\}}
exp(-i2\varphi x_2).
\end{equation}
Hence, we have the relations
\[
\left(\phi^1_I\right)^{+}\neq \left(\phi^1_I\right)^{-}\neq
\left(\phi^2_I\right)^{+}\neq \left(\phi^2_I\right)^{-},\ \
\left(\phi^1_{II}\right)^{+}=\left(\phi^1_{II}\right)^{-}\neq
\left(\phi^2_{II}\right)^{+}=\left(\phi^2_{II}\right)^{-} .
\]
When $y=1$ our generalized Kronig-Penney relations (26) and (33) reduce to
the corresponding Kronig-Penney relations given in [18].

In conclusion we can say that the differences between the wave functions (13)
and (14) and between the wave functions (27) and (28), which finally depend
on the choice of unit cell and the matching conditions, should be taken into
account when we compute surface and interface electronic states. In other
words, varying the position of the barrier-well interface within an unit
cell, we will influence the conditions for the existence of surface and
interface states. For the two potentials (1) and (25) our generalized
Kronig-Penney relations (12) and (26), and the particular case (33), are
different, as it is in the original Kronig-Penney relations, and the
corresponding Bloch wave functions for these two potentials are also
different.

\newpage
%\vskip 1cm
{\bf REFERENCES}
\vskip 0.5cm
[1] Bastard G., {\it Wave Mechanics Applied to Semiconductor
Heterostructures}, (Les Editions \hspace* {1 cm} de Physique, Les Ulis)
1988.

[2] Bloss W. L., {\it Phys. Rev. B}, {\bf 44}(1991) 8035.

[3] Shen M. R., Shen W. Z. and Li Z. Y., {\it phys. stat. sol}. (b) {\bf
177}(1993) K71.

[4] Ste\'slicka M., in {\it Progr. Surf. Sci}., vol. {\bf 5}, edited by S. G.
Davison (Pergamon Press Ltd) \hspace* {1 cm} 1975, 157.

[5] St\c{e}\'slicka M., Kucharczyk R. and Glasser M. L., {\it Phys. Rev. B}
{\bf 42}(1990) 1458.

[6] St\c{e}\'slicka M., {\it Progr. Surf. Sci}., {\bf 48} (1995) 207

[7] Kandilarov B. D., Tashkova M., Petrova P. and Detcheva V., {\it
phys. stat. sol} (b) {\bf 86} (1978) \hspace* {1 cm} 425.

[8] Kandilarov B. D., Tashkova M., and Petrova P., {\it
phys. stat. sol} (b) {\bf 87} (1978) 361.

[9] Kandilarov B. D., Detcheva V. and Primatarova M. T., {\it J. Phys. C},
{\bf 12} (1979) 3401.

[10] Carpena P., Gasparian V. and Ortuno >, {\it European Physical Journal
B}, {\bf 8} (1999) 635.

[11] Kronig R. de L. and Penney W. G., {\it Proc. Roy. Soc. A}, {\bf 130}
(1931) 499.

[12] Gubanov A. I. and Sharapov B. N., {\it Izvestiya LETI}, {\bf 64} (1969)
52

[13] Metodieva M. and Detcheva V., {\it Comp. rend. Acad. bulg. Sci}., {\bf
39} (1986) 35.

[14] Bejenari I., Malkova N and Kanster V. G., {\it phys. stat. sol.} (b)
{\bf 223} (2001) 785.

[15] Harrison W. A., {\it Phys. Rev.}, {\bf 123} (1961) 85.

[16] Ben Daniel D. J and Duke C. B., {\it Phys. Rev.}, {\bf 152} (1966) 683.

[17] Bastard G., {\it Phys. Rev. B}, {\bf 24} (1981) 5693.

[18] Haug A., {\it Theoretical Solid State Physics}, vol.1, Chp.II, (Oxford)
1972.

\newpage
Fig.1: One-dimensional Kronig-Penney model with rectangular barriers and
lattice constant $(a+b)$: $b$-the width of the barriers, $V$-the height of
the barriers: (a) unit cell of KP-1 type: "region II- region I"
(well-barrier); (b) unit cell of KP-2 type: "region I- region II"
(barrier-well).
%\newpage
\vskip 0.5cm
Fig.2: One-dimensional Kronig-Penney model with rectangular wells and
lattice constant $(a+b)$: $b$-the width of the wells, $V$-the depth of
the wells: (a) unit cell of KP-1a type: "region II- region I"
(well-barrier); (b) unit cell of KP-2a type: "region I- region II"
(barrier-well).

 \end{document}